\documentclass[floatfix,aps,twocolumn,superscriptaddress,prb]{revtex4-2}

\usepackage{graphicx}
\usepackage{dcolumn}
\usepackage{bm}
\usepackage[T1]{fontenc}
\usepackage[latin9]{inputenc}
\usepackage[linkcolor=blue,colorlinks=true]{hyperref}
\usepackage{textcomp}
\usepackage{amsmath}
\usepackage{mathrsfs}
\usepackage{graphicx}
\usepackage{amssymb}
\usepackage{units}
\usepackage{xcolor}
\usepackage{float}

\newcommand{\ivan}[1]{\textcolor{black}{#1}}

\definecolor{emerald}{rgb}{0.31, 0.78, 0.47}

\begin{document}

\title{First order Quantum Hall to Wigner crystal phase transition on a triangular lattice: an iDMRG study}

\author{Gleb Fedorovich}
\email{gleb.fedorovich@ugent.be}
\affiliation{
Institute of Quantum Electronics ETH Zurich, CH-8093 Zurich, Switzerland}

\affiliation{Department of Physics and Astronomy, Ghent University, Krijgslaan 281, 9000 Gent, Belgium}

\author{Clemens Kuhlenkamp}
\affiliation{
Institute of Quantum Electronics ETH Zurich, CH-8093 Zurich, Switzerland}

\author{Atac Imamoglu}
\affiliation{
Institute of Quantum Electronics ETH Zurich, CH-8093 Zurich, Switzerland}

\author{Ivan Amelio}
\affiliation{
Institute of Quantum Electronics ETH Zurich, CH-8093 Zurich, Switzerland}
\affiliation{Center for Nonlinear Phenomena and Complex Systems,
Universit{\'e} Libre de Bruxelles, CP 231, Campus Plaine, B-1050 Brussels, Belgium}
\affiliation{International Solvay Institutes, Brussels, Belgium}

\begin{abstract}
In this work we study a system of interacting fermions on a triangular lattice in the presence of an external magnetic field. We \ivan{assume electrons are spin-polarized} 
and fix a density of one third, with one unit of magnetic flux per particle.
The infinite density matrix renormalization group algorithm is used to compute the ground state of this generalized Fermi-Hubbard model.
Increasing the strength of the nearest-neighbor repulsion, we find a first order transition between an Integer Quantum Hall phase and a crystalline, generalized Wigner crystal state. 
The first-order nature of the phase transition is  consistent with a Ginzburg-Landau argument.
We expect our results to be relevant for 
moir\'e heterostructures of two-dimensional materials.
\end{abstract}

\date{\today}
\maketitle

\section{Introduction}

Two-dimensional materials have recently emerged as an extremely suitable platform to investigate many-body physics~\cite{Andrei2021,Mak2022}. 
The small electrostatic screening as compared to bulk materials together with the engineering of weakly dispersive moir\'e bands ensure access to the regime of strong correlations~\cite{wu2018hubbard,Tang2020simulation,shimazaki2020strongly}.
The possibility of stacking different heterostructures allows to have a precise control of the system parameters.
In just a few years many groups have been able to experimentally realize many exotic phases of matter, including superconductivity in twisted bilayer graphene~\cite{cao2018super},
(fractional) Chern 
insulators~\cite{Nuckolls2020,Xie2021},
excitonic insulators~\cite{ma2021strongly,gu2022dipolar,Sun2021evidence}, anomalous quantum Hall effect~\cite{Li2021quantum, Park2023}, and different kinds of charge density waves~\cite{Regan2020mott,Xu2020correlated,smolenski2021signatures,Zhou2021bilayer,shimazaki2021optical,polovnikov2022coulombcorrelated,tsui2023direct}. 
In particular, in the zoo of possible charge density waves we can distinguish the Wigner crystals in the strict sense of the term, where {\em continuous} translation symmetry is spontaneously broken~\cite{smolenski2021signatures,Zhou2021bilayer,shimazaki2021optical,tsui2023direct}, and generalized Wigner crystals, when  a charge density wave is formed on an underlying (e.g. moir\'e) lattice, so that only a {\em discrete} translation
symmetry is spontaneously broken~\cite{Regan2020mott,Xu2020correlated,polovnikov2022coulombcorrelated}.

Two-dimensional systems are also special for topological reasons, since a Chern number can be defined in even dimensions and a non-trivial braiding of anyons is possible only in 2D.
When applying an external magnetic field to a 2D electron gas, Integer and Fractional Quantum Hall states can be created~\cite{klitzing1980,tsui1982}.
The competition between Wigner crystals and Quantum Hall states as a function of density and magnetic fields is a fundamental and well studied topic in the field~\cite{levesque1984,koulakov1996charge,Haldane2000,zuo2020interplay,meng2020thermal,villegasrosales2021competition,yang2023cascade}.
~\ivan{In particular, in continuum systems it is believed that the liquid-crystal transition occurs via intermediate continuous transitions between bubble and stripe phases~\cite{spivak2004,falson2022competing}.
However, the situation when an underlying lattice  is present has received much less attention, and the transition between
the Chern insulator and the generalized Wigner crystal, deserves  investigation,  as well as the possibility of intermediate phases.
In a similar spirit, recent numerical studies have revealed an intermediate bond-ordered phase in the context of the interaction-driven nematic crystal to quantum anomalous Hall transition
 in quadratic band touching systems~\cite{Sur2018,Zeng2018}.
Another open question, relevant for experiments, concerns the parameter regimes at which the transition occurs in moir\'e TMD heterostructures.
Moreover, this setting} is well suited to be studied with tensor network techniques, such as the Density Matrix Renormalization Group (DMRG) algorithm~\cite{White1992Nov, Schollwock2005Apr}. A very recent DMRG study in the absence of an external magnetic field demonstrated a first order quantum phase transition between a Fermi sea and a generalized Wigner crystal, with 120$^{\circ}$
Neel spin order~\cite{zhou2024quantum}.
The spin order in the crystalline phase was also studied in  \cite{moralesduran2023,seifert2024spinpolarons,biborski2024chargespinpropertiesgeneralized}.
In \cite{kuhlenkamp2022tunable}, instead, two of us demonstrated, using the iDMRG technique, the emergence of chiral spin liquid physics on top of a Mott insulating state exposed to an external magnetic field \ivan{and found evidence for continuous phase transitions} (see also \cite{KuhlenkampThesis,divic2024chiralspinliquidquantum}).
\ivan{DMRG} 
was also applied to study Wigner crystallization on a ladder~\cite{kiely2023continuous}  and its competition with superconducting pairing~\cite{clay2023density}.

\begin{figure*}
    \centering
    \includegraphics[width=2\columnwidth]{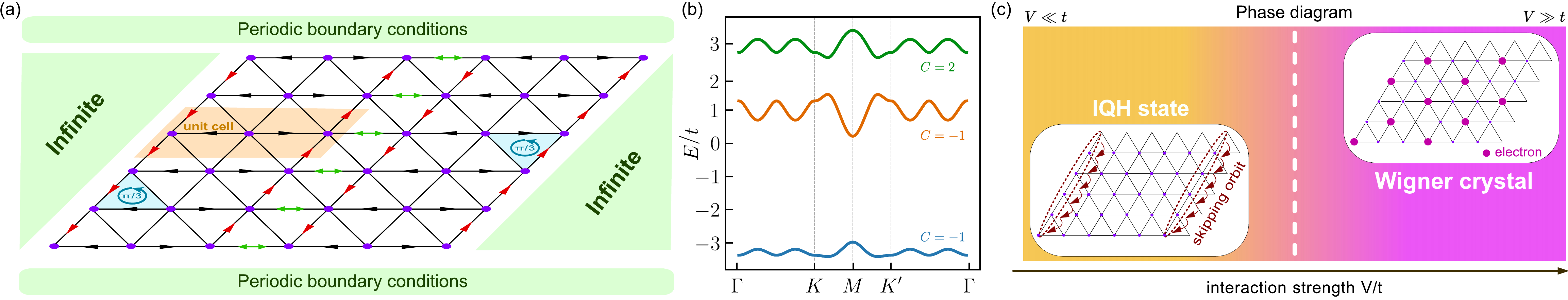}
    \caption{
    (a)  Triangular lattice in a uniform magnetic field and on an infinite cylinder geometry, with periodic boundary conditions along $y$. Black, green and red arrows 
    illustrate a computationally convenient gauge choice and
    correspond to $\pi/3, \pi, 2\pi/3$ hopping phases, respectively, resulting in a $\Phi = \pi/3$ flux per triangle. (b) 
    \ivan{
    Single-particle bands of the triangular lattice in a magnetic field, for a representative path in quasi-momentum space. The  $\Phi = \pi/3$ flux
    gives rise to three bands with Chern number $C = -1,-1,2$, in order of increasing energy.
    (c)}
Schematic phase diagram illustrating the transition from the IQH  to the WC state for increasing repulsion $V$ between the fermions at density $n_e=1/3$.}
    \label{fig1}
\end{figure*}

In this work we address the formation of a  generalized Wigner crystal (WC) state in a triangular lattice, and the competition of this ordered phase with the Integer Quantum Hall (IQH) liquid.
We fix the density to one fermion every three sites to ensure the commensurability of the density wave, and choose the magnetic field so to have one unit of flux per particle. \ivan{Due to the strong magnetic field we consider spin-polarized fermions, and model screened Coulomb interactions in the system by taking into account nearest-neighbor (NN) repulsion.} 
We then obtain the ground state numerically by means of infinite density matrix renormalization group (iDMRG) method.
More specifically, in Section \ref{sec:model}
we introduce the model and the iDMRG algorithm.
Numerical results for the ground state energy, entanglement entropy, density modulation and differential overlap are reported in Section \ref{sec:results}.
Following the numerical evidence of a first order phase transition between the IQH and WC phases when the strength of electron-electron repulsion is increased, in Section 
\ref{sec:GL}
we support this observation with a symmetry argument based on the Ginzburg-Landau free energy approach to phase transitions. The experimental observability of the phase transition from IQH to WC state in twisted TMD heterostructures is discussed in Sec.~\ref{app:exp}.
We then present conclusions and the outlook in Section \ref{sec:conclusions}.



\section{Model and methods}
\label{sec:model}

In the following, we consider an extended Hubbard model of spinless fermions on a triangular lattice, with an external magnetic field and finite range interactions.   The Hamiltonian   can be written in  real space as
\begin{equation}
\label{spinless_H_real}
\hat{H}_{xy} = -\sum_{\langle i,j \rangle} (t_{ij} c^\dagger_{i}c_{j} + h.c.) + V\sum_{\langle i,j \rangle}n_{i}n_{j},
\end{equation}
where $c^\dagger_{{i}},c_{i}$ are creation/annihilation fermionic operators acting on  site $i$, with $i = (x, y)$  belonging to a two-dimensional triangular lattice. Here $t_{ij} = te^{i\phi_{ij}}$ stands for the complex hopping amplitude between two sites $i$ and $j$, the sums are restricted to nearest neighbours pairs $\langle i,j \rangle$, and $V$ is the strength of the repulsive interaction between electrons in neighbouring sites. 
Importantly, as we demonstrate below in Section \ref{app:exp},
our model and its critical regime can be realized in TMD heterostructures, such as in MoSe$_2$/WSe$_2$ twisted bilayers or WSe$_2$ with a proximal hBN twisted bilayer, for experimentally accessible parameters. In particular, we relate the ratio $V/t$ to the twist angle and gate distance, and report the required values of the external magnetic field.

In order to target at the same time the Integer Quantum Hall (IQH) and crystalline states, we set the electron density and the magnetic flux per triangle to $n_e = 1/3$ and $\Phi = \pi/3$, respectively. As a result, the filling factor $\nu$ 
\ivan{equals unity}:
\begin{equation}
    \nu = 2\pi \cdot \frac{n_e}{\Phi_{\text{u.c.}}} 
    = 1,
\end{equation}
where $\Phi_{\text{u.c.}} = 2\Phi$ stands for the flux per unit cell, the unit cell consisting of two triangles. 
While with iDMRG one could in principle study also fractional fillings,
the IQH phase typically has the largest many-body gap and is quite robust against the curvature of the Chern bands.
We also mention the recent iDMRG study of the stabilization by an external magnetic field of a chiral spin liquid in a spinful Fermi-Hubbard
on a triangular lattice~\cite{kuhlenkamp2022tunable}.

In the numerical treatment below, we will adopt the infinite cylinder geometry sketched in Fig.~\ref{fig1}a), where a finite number of sites and periodic boundary conditions are considered along the $y$-axis, while translational invariance allows to treat the $x$ direction as infinite.
There are several gauge choices for the hopping phases $\phi_{ij}$  that yield a uniform magnetic flux of $\Phi = \pi/3$ per triangle. The one used in our calculations is illustrated in Fig.~\ref{fig1}a), where  the arrows are color-coded so that black,  green
and red are associated with $\pi/3, \pi$ and $2\pi/3$ hopping phases, respectively. 
Within this gauge, the magnetic unit cell extends for three unit cells in the $x$ direction. 
\ivan{The single-particle bands are sketched in Fig.~\ref{fig1}.
b), and are topological. Their Chern numbers are, going from low to high energy, $-1,-1,2$.}

The schematic phase diagram as a function of the interaction strength $V/t$ is shown in Fig.~\ref{fig1}c).
While the IQH phase is expected to occur for zero and weak interactions between the fermions, for strong repulsion $V \gg t$ we expect the formation of a charge density wave, which we refer to as a Wigner crystal (WC). Notice that sometimes the name extended or generalized Wigner crystal is used in the literature~\cite{Regan2020mott,Li2021}. 
With the electron filling set to $n_e=1/3$, the crystal superlattice is commensurate with the microscopic lattice and is shown on the right side of Fig.~\ref{fig1}c).
The main goal of this work is to investigate the phase transition between the IQH and WC phases.
 
\ivan{To study the system numerically using iDMRG, it is advantageous to make use of the $U(1)$ charge and $\mathbb{Z}_{L_y}$ translational symmetry. To do this we implement the Hamiltonian in a mixed real and momentum space representation, following~\cite{Motruk2016Apr}. After Fourier transforming in $y-$direction, the Hamiltonian reads:}

\begin{equation}
\begin{split}
    \hat{H}_{xk} = \sum_{x,k_y} \left(\tilde{t}_{1x} c_{x+1,k_y}^\dagger c_{x,k_y} + h.c.\right) + \sum_{x,k_y} \tilde{t}_{2x}
    c_{x,k_y}^\dagger c_{x,k_y} + \\
    + \frac{V}{L_y}\sum_{x, q_y}\left(\cos{q_y\cdot}n_{x,q_y}n_{x, -q_y} + \left(1 + e^{-iq_y}\right)n_{x+1,q_y}n_{x, -q_y} \right),
\end{split}
\end{equation}
where $n_{x,q_y} = \sum_{k_y}c^\dagger_{x,k_y+q_y}c_{x,k_y}$
and momenta run over
$k_y = \frac{2\pi}{L_y}k$, with $k=0,1,...,L_y-1$. The kinetic part contains the hopping terms $\tilde{t}_{1x}$ and $\tilde{t}_{2x}$, which depend on the position of $x$ within the magnetic unit cell, namely
\begin{equation}
    \begin{split}
        \tilde{t}_{1x} = -t(e^{i\Phi(2x+1)} + e^{-ik_y}), \\        
        \tilde{t}_{2x} = -2t\cos(k_y+2\Phi x).
    \end{split}
\end{equation}

In the following, we compute the ground state of the Hamiltonian $\hat{H}_{x,k_y}$ using the iDMRG implementation from the  TenPy library \cite{tenpy}. 
This variational approach relies on translational invariance along $x$ and addresses  matrix product states living on an infinite system by introducing a computational unit cell over which the tensors representing the variational state are defined.
For the size of the computational unit cell we fix  $L_x = 6$ and vary $L_y = 9, 12$ (where $L_y$ is also the actual size of the system along $y$).
\ivan{
We verified for a selected number of parameters that the ground state always occurs in the sector of zero total momentum~\footnote{
A few other total momentum sectors display a low energy ground state, yet slightly higher in energy than the zero momentum one. We attribute this to the structure of the Chern and Wigner quasi-particle bands, rather than to some finite-size manifestation of non-Abelian topological degeneracies~\cite{Szasz2020chiral}.
}, and its energy and properties agree with the ones obtained from the  real space computation, without momentum constraint.
Overall,} two Abelian mutually commuting symmetries are imposed in the calculation, namely we fix the number of fermions and select the sector of zero total momentum along $y$. To reveal the competition between different phases that emerge in the system, we calculate  the energy,  the density and crystalline order parameter, the momentum-resolved entanglement spectrum,  the fidelity susceptibility
\ivan{and the correlation length}. 


\section{Results}
\label{sec:results}

In this Section we report our numerical results obtained via the iDMRG method.
We start by characterizing the competition between two phases of matter, the Wigner Crystal emerging at large interactions, and an Integer Quantum Hall liquid, dominated by the single-particle term of the Hamiltonian.
We then focus on the singularity in the energy slope  occurring around $V/t \simeq 3.7$, as well as on the jump of the crystal order parameter and the peak in the fidelity susceptibility,  hinting at a first order  Wigner-Quantum Hall phase transition.

\subsection{Wigner crystal phase}
\label{chapter_WC}

Let's start considering the right side of the phase diagram, $V\gg t$.
When the repulsion between electrons dominates over the kinetic hopping term, we expect the formation of a charge density wave,
which minimizes the repulsive potential energy.
In the commensurate case of one electron every three lattice sites $n_e=1/3$, 
one has the classical configuration in which
electrons fill a triangular superlattice with  unit Bravais vectors of length $\sqrt{3}$, as depicted in Fig.~1b). 
At the quantum level, the translational symmetry can be spontaneously broken in an infinite system, and the classical crystalline configurations will be smoothened by quantum fluctuations stemming from the kinetic term of the Hamiltonian. 

\begin{figure}
    \centering
    \includegraphics[width=\columnwidth]{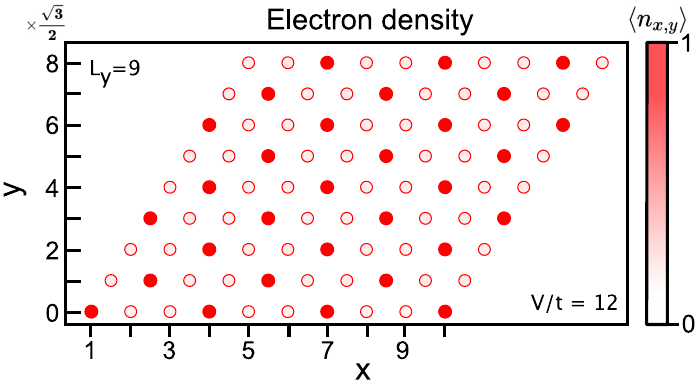}
    \caption{
    Ground state  electron density in the large interaction regime
    $V/t=12$,  displaying the spontaneous formation of a Wigner crystal at density $1/3$.    This result has been obtained via iDMRG, with $L_y=9$ and bond dimension $\chi = 2000$.}
    \label{fig2}
\end{figure}

One key feature of iDMRG is that it supports spontaneous symmetry breaking~\cite{McCulloch2008Apr}.
As a result, we plot in Fig.~\ref{fig2} the  density of the iDMRG ground state, color-coded in such a way that white and red correspond to small ($n_{x,y} \approx 0$ - empty site) and large ($n_{x,y} \approx 1$ - occupied site) densities, respectively. 
Depending on the initial condition, different  crystalline configurations are selected, which are all equivalent modulo a discrete translation.
In Fig.~\ref{fig2}, a cylinder length of $L_y = 9$ and an interaction strength of $V/t = 12$
were used.

\subsection{Integer Quantum Hall phase }\label{chapter_IQH}

\begin{figure}
    \centering
    \includegraphics[width=\columnwidth]{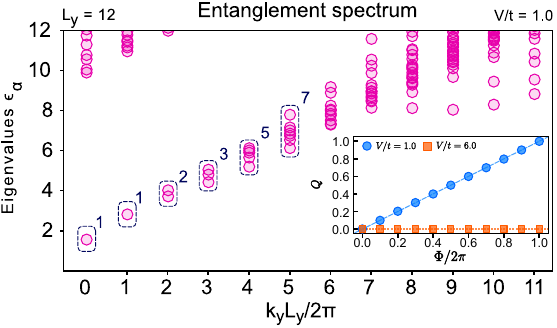}
    \caption{
    iDMRG momentum-resolved entanglement spectrum $\epsilon_\alpha$ of the ground state for   $L_y = 12$ as a function of normalized momentum $k_y \cdot L_y/2\pi$, for interaction strength $V/t = 1.0$. The degeneracy of the lowest entanglement energies $\{1,1,2,3,5,7,...\}$ together with their linear dispersion   indicate the 
    presence of a chiral edge mode.
    This supports the \ivan{existence of an IQH state}. The bond dimension used in this computation was $\chi = 3000$. \ivan{Inset: charge $Q$ pumped into the system via flux insertion. In the Integer Quantum Hall phase a unit of charge is inserted for each quantum of flux (blue circles). In the WC phase (orange squares), instead, no Hall conductance is observed.}
    }
    \label{fig3}
\end{figure}

In the opposite regime of small $V/t$,
the kinetic term of the Hamiltonian dominates over interactions, and the ground state is an Integer Quantum Hall fluid. 
Below we confirm this expectation by numerically computing the momentum resolved entanglement spectrum~\cite{Pollmann2012Sep}
and showing that it carries the hallmark of a chiral edge mode. 

To this end, we first cut the tensor network into two halves and leverage 
the translation symmetry along the $y$-direction. 
An entanglement energy level ${\epsilon_\alpha}$ is linked to the Schmidt decomposition values $\Lambda_\alpha$~\cite{Cincio2013Feb, Li2008Jul} of the cut bond via the relation $\Lambda_\alpha = e^{-\epsilon_\alpha/2}$, with the Schmidt decomposition being performed separately in each momentum sector.
For
$V/t=1$ the momentum-resolved entanglement spectrum is shown in Fig.~\ref{fig3} and
the degeneracy of the entanglement energies follows the pattern $(1,1,2,3,5,7,...)$.
This is the expected result for the energy spectrum of a chiral Luttinger liquid, which describes the neutral excitations of a IQH edge~\cite{Haldane1981Jul, BibEntry2023Oct}.
Notice that the presence of an entanglement gap $\sim 10$, due to the interaction with the bulk of the system, together with the diameter $L_y$ of the cylinder determine the number of chiral excitations which can be resolved. 

\ivan{While the counting is fixed by the chiral Luttinger liquid, the topological charge response of the system cannot be immediately inferred. To unveil the Hall response, we computed the amount of electric charged pumped under flux insertion}
~\cite{Zaletel2014,Grushin2015} in the inset of Fig.~\ref{fig3} (blue circles). This displays one unit of charge \ivan{is transferred} per unit of flux, confirming the 
\ivan{existence} of {\em Integer} Quantum Hall physics,
in agreement with the single-particle  Chern number of the lowest band via the Thouless-Kohmoto-Nightingale-den Nijs
formula~\cite{thouless1982}.
The same calculation confirms the absence of \ivan{a quantized Hall response} 
in the WC phase (orange squares).


\ivan{\subsection{Phase Transition}}
\ivan{Here we study the nature of the IQH to generalized WC transition, by analyzing the ground state energy, studying the discontinuity of the WC's order parameter, the correlation length, as well as the ground state overlap in the two different phases.}
\subsubsection{Energy}

\begin{figure}[t]
    \centering
    \includegraphics[width=\columnwidth]{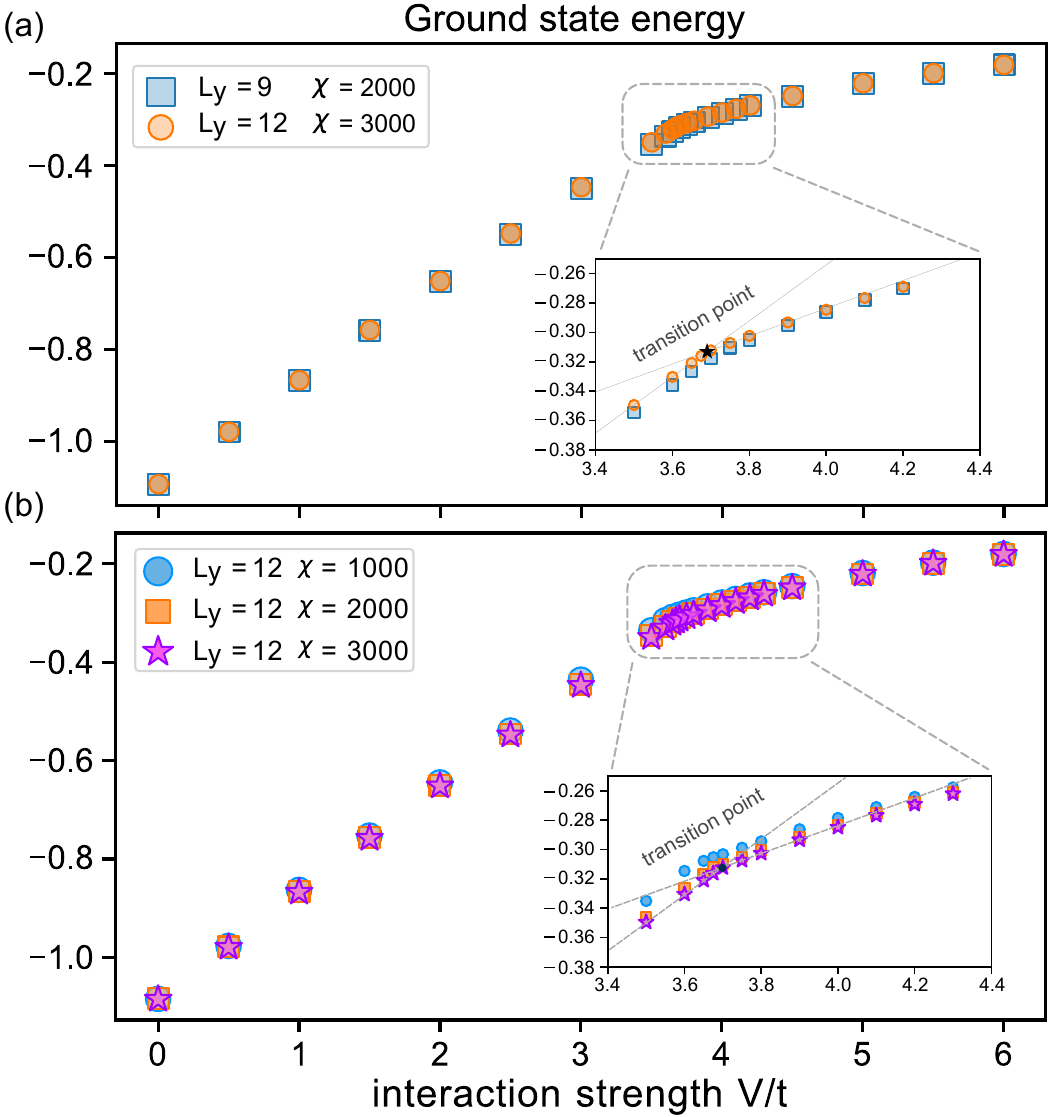}
    \caption{
    (a) Ground state energy per  site as a function of interaction strength $V/t$ for $L_y=12$ (orange circles)
    and $L_y=12$
    (blue squares).
    In the inset we zoom in on the transition point, and to highlight the derivative singularity we draw the linear fit lines above and below the transition (indicated by the star). 
    (b) Ground state energy for increasing bond dimension $\chi=1000,2000,3000$.}
    \label{fig:energy}
\end{figure}

In Fig.~\ref{fig:energy}a)
we plot the energy
as a function of $V/t$
for two cylinder diameters 
$L_y=9$ and $L_y=12$.
As highlighted in the inset, which zooms in on the area around the phase transition point, the convergence with system size is excellent and \ivan{further improves upon} 
going deeper into the crystalline phase, as one would expect.
The linear fits (gray dashed lines) below and above the phase transition
point (denoted by the black star)
clearly reveal the
singularity in the energy, more precisely, the discontinuity of its derivative.

We recall that, for zero temperature quantum phase transitions driven by a continuous change of the Hamiltonian, the ground state energy is always continuous through the transition.
(The analogue in classical phase transitions at finite temperature is the free energy, which is always continuous, while the energy and other order parameters can be discontinuous.)
 In Fig.~\ref{fig:energy}b), we instead fix
$L_y=12$, and
demonstrate the convergence of the energy with increasing bond dimensions, namely $\chi=1000,2000,3000$. \ivan{The truncation error of the last iDMRG step is of the order of $10^{-4}$ at the transition point, and becomes up to two orders lower in both the IQH and WC phases (not shown).}

\subsubsection{Order parameter}
In Fig.\ref{fig2}
we showed that the density in the ground state matches the triangular superlattice pattern expected for a WC.
In order to be more quantitative and address values of $V/t$ close to the phase transition, 
we define the order parameter 
$\eta$
as the Fourier transform of the real space electron density $n_{\mathbf{r}}$:
\begin{equation}
   \eta
   =
   n_{\mathbf{G}_1} = \sum\limits_{\mathbf{r}}  n_{\mathbf{r}}   \cdot e^{i\mathbf{G}_1 \mathbf{r}},
\end{equation}
where the sum over $\mathbf{r}$ includes all the sites in the iDMRG computational block. Here $\mathbf{G}_1 =(2\pi/3, 2\pi/\sqrt{3})$ is one arbitrary generator of the reciprocal lattice associated with the WC superlattice. 
\ivan{(Notice that for the specific case of the triangular lattice at one-third filling considered here, $\mathbf{G}_1$ is also equivalent to the $\mathbf{K}$ point of the microscopic lattice.)}
Therefore, for a triangular WC state we expect the order parameter to be nonzero, while it will be zero for a liquid state which does not break translational symmetry.

\begin{figure}
    \centering
    \includegraphics[width=\columnwidth]{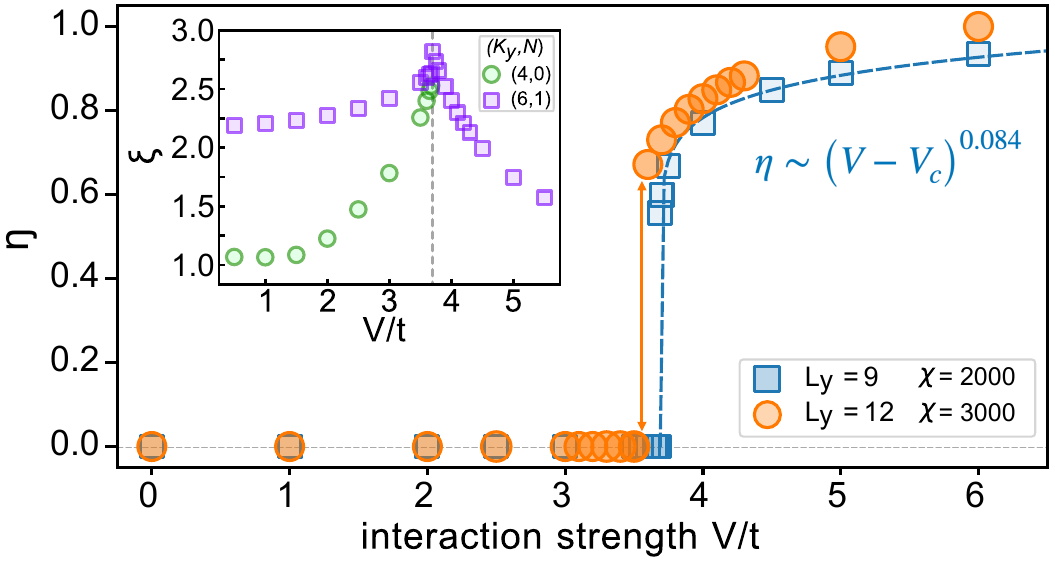}
    \caption{
    Order parameter $\eta$, corresponding to the strength of the WC density modulation, as a function of  $V/t$, for both  $L_y = 12$ (orange circles) and $L_y=9$ (blue squares). The blue dashed line represents \ivan{a fit to} 
    the $L_y = 9$ data with a power-law, which yields a very small exponent (hinting for a jump in the order parameter, which is even more clear for $L_y=12$). \ivan{The inset demonstrates correlation lengths $\xi$ obtained in two different symmetry sectors, labeled by the total momentum $K_y$ (in units of $2\pi/L_y$) and by the fermion number $N$. 
For the $(K_y,N)=(4,0)$ sector (green circles), $\xi$ diverges in the WC phase, due to the spontaneous symmetry breaking. Numerical parameters: $L_y =12, \chi = 3000$.}}
    \label{fig:OP}
\end{figure}

We report in Fig.\ref{fig:OP} the normalised order parameter as a function of the interaction strength $V/t$ for  $L_y = 9$ (blue squares) and  $L_y = 12$ (orange circle).
While for $L_y = 9$ the singularity in $\eta$ could be still fit with a rather small power law 
$\eta(V>V_c) \sim (V-V_c)^{0.084}$,
a jump clearly opens up for $L_y=12$, suggesting a genuine first order transition.

\subsubsection{Correlation length}

\ivan{As a next numerical result, it is also interesting to inspect the correlation length of the system. 
In gapped systems, the  correlation function at equal times of a local operator $O_j$ typically decays exponentially with a spatial rate determined by the correlation length $\xi$. 
In our case, 
we have two conserved quantities, total momentum along $y$ and fermion number, and 
different $\xi$'s are obtained depending on the charge of operator $O_j$ with respect to these two symmetries.
In the iDMRG method, $\xi$ is easily determined from the largest eigenvalue of the transfer matrix in the proper charge sector.
 In the inset of Fig.~\ref{fig:OP}, we report the  correlation length in the charge sectors
$(K_y,N)=(4,0)$ and $(K_y,N)=(6,1)$, for which we find the largest $\xi$'s.
The correlation length in the sector $(K_y,N)=(4,0)$ corresponds to the green circles and diverges above the transition, as expected from $k_y=4 \frac{2\pi}{L_y}$ being the wavevector of the charge density wave.
The other sector where $\xi$ is particularly large is $(K_y,N)=(6,1)$, in violet squares, and relates to a single-particle Green's function. It is not divergent in the WC phase, but increases and makes a jump around the transition point. These results strengthen our picture of a first order transition with sizable quantum fluctuations.
}

\subsubsection{Differential overlap}

Since iDMRG provides us with the variational wavefunction of the GS, we can also compute the overlap between GS's at different $V/t$, which is a very sensitive quantity to the presence of a quantum phase transition~\cite{zanardi2006ground}.

\begin{figure}
    \centering
    \includegraphics[width=\columnwidth]{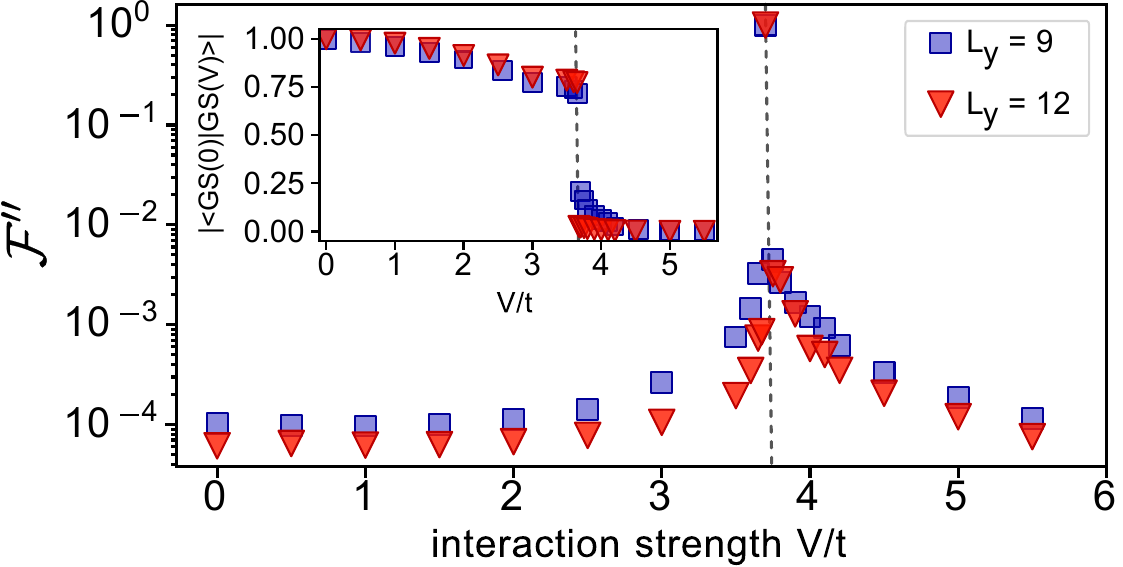}
    \caption{
The fidelity susceptibility $\mathcal{F}''(V)$,  defined in Eq.~\ref{eq:diff_overlap},
is plotted as a function of the interaction strength $V/t$, for  $L_y = 9$ (blue squares) and $L_y = 12$ (red triangles). We used bond dimensions of $\chi = 2000$ and $\chi = 3000$, respectively. A $\delta$-like peak is clearly visible at the phase transition point $V_c \approx 3.7.$
\ivan{The inset  shows the overlap  between the ground state at $V \neq 0$ with the non-interacting IQH state as a function of interaction strength.}
}
\label{fig:overlaps_new}
\end{figure}

More precisely, we
introduce the  fidelity susceptibility~\cite{wang2015fidelity}
\begin{equation}
\mathcal{F}''(V) =
\left.
-\frac{\partial^2}{\partial\epsilon^2}
|\langle GS(V) | GS(V+ \epsilon)| \rangle_{\rm cell} 
\right|_{\epsilon=0},
\label{eq:diff_overlap}
\end{equation}
where the overlap of the ground states here is computed from the dominant transfer matrix eigenvalue, and represents the contribution from each computational  cell. (Strictly speaking, the overlap between states would be the product of the cell contributions, which is vanishing for an infinite system.) In Fig.~\ref{fig:overlaps_new} we plot $\mathcal{F}''(V)$, where 
the derivative is evaluated by discretizing the interaction strength step.

While for a second phase transition one would expect a power law divergence of this quantity~\cite{zanardi2006ground}, here we observe a Dirac delta singularity at the transition point, confirming the fact that the transition is first order.
Nonetheless, the moderate increase of $\mathcal{F}''(V)$
close to the transition point, which is in correspondence with the decrease of $\eta$ in Fig.~\ref{fig:OP}, suggests that the transition is {\em weakly} first order, in the sense that  some strong fluctuations are  present in the proximity of the transition.
\ivan{A final important remark is that  no other peaks are visible in the fidelity susceptibility, ruling out the possibility of intermediate phases (at least for the system sizes considered here).}

\section{Ginzburg-Landau argument for a first order transition}
\label{sec:GL}

Above we have numerically demonstrated that, for increasing $V/t$, the IQH fluid crystallizes into a WC via a first order phase transition. Here, we want to recall an argument based on the Ginzburg-Landau formalism, stating that crystallization into a {\em triangular} charge density wave occurs generically via a {\em first} order transition~\cite{Tinkham1975IntroductionTS}.
While this argument is  proven 
for generic symmetry groups
in the classical Landau-Lifshitz book on statistical mechanics~\cite{landau2013statistical},
we think it will be helpful to reformulate it here in our specific case. 

 The free energy, as a function of the order parameter, is the central object of the Ginzburg-Landau approach.
 For a zero temperature quantum phase transition,
 the free energy corresponds to the energy expressed as a function of some coarse grained degrees of freedom, which are sometimes referred to as reaction coordinates or order parameters. At finite temperature, also the entropy contributes to the free energy.
 The free energy functional should be invariant under all the symmetry transformations of the physical system, and this  constraints  the terms allowed in the general expression of the  functional.

\begin{figure*}[t]
    \centering
    \includegraphics[width=2\columnwidth]{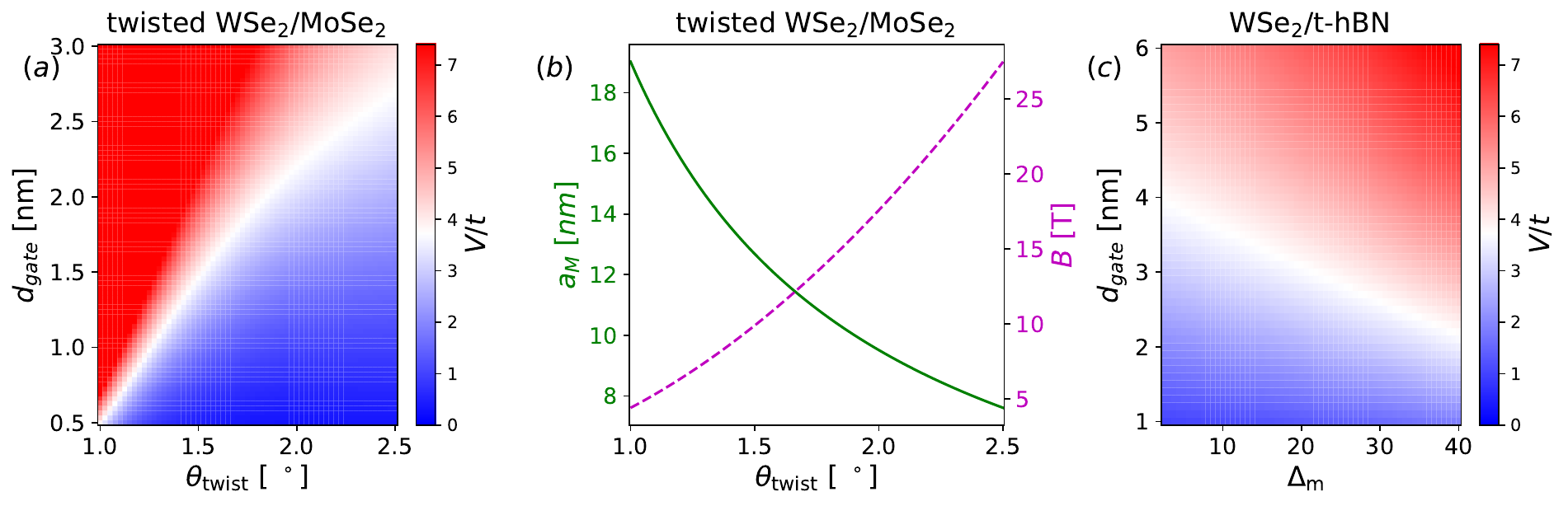}
    \caption{
    (a)
    Ratio $V/t$ of nearest-neighbor Hubbard interaction to hopping, calculated from Wannier orbitals of a MoSe$_2$/WSe$_2$ bilayer as a function of twist angle and gate distance. The red and blue colors correspond to the WC and IQH regions, respectively, while the white area is the critical region.
    (b) The moir\'e unit length $a_M$ is plotted as a function of $\theta_{\rm twist}$ (solid green line, left axis). The magnetic field to achieve a flux of $\pi/3$ per triangle is also shown (dashed magenta, right axis).
    (c) The ratio $V/t$ is now shown for a WSe$_2$ monolayer with an external potential induced by a proximal twisted hBN bilayer. On the horizontal axis, the depth of the electrostatic potential $\Delta_{\rm m}$ can be tuned by inserting a hBN spacer.
    }
\label{fig:WannierHubbard}
\end{figure*}

More precisely, in our case we want to express the Ginzburg-Landau free energy as a functional of the density  
\begin{equation}
    n(\mathbf{r})
    = n_0 +
    \eta(\mathbf{r}) = n_0 + \sum_\mathbf{G} 
    e^{i\mathbf{G}\mathbf{r}}
    \eta_{\mathbf{G}},
\end{equation}
where the sum runs over the reciprocal space of the crystal and $\eta$ represents density variations with respect to the average density $n_0=1/3$. 
The Ginzburg-Landau functional $F$ can then be
formally expanded in the Taylor series
\begin{equation}
    F[n(\mathbf{r})]
    =
    F[0] 
    +
    \sum_{n=1}^{+\infty} \frac{1}{n!}
    \sum_{\mathbf{r}_1...\mathbf{r}_n} 
    f^{(n)}_{\mathbf{r}_1...\mathbf{r}_n}
    \eta(\mathbf{r}_1) ... \eta(\mathbf{r}_n),
\end{equation}
where  the shortcut
$
f^{(n)}_{\mathbf{r}_1...\mathbf{r}_n}
=
\left.
    \frac{\delta^{(n)}F}{\delta\eta(\mathbf{r}_1)...\delta\eta(\mathbf{r}_n)}
    \right|_{\eta=0}$
was introduced for the functional derivatives.
Fourier transforming as 
$f^{(n)}_{\mathbf{G}_1...\mathbf{G}_n}
=
\sum_{\mathbf{r}_1...\mathbf{r}_n}
e^{i\mathbf{G}_1\mathbf{r}_1+...+i\mathbf{G}_n\mathbf{r}_n}
f^{(n)}_{\mathbf{r}_1...\mathbf{r}_n}$,
we get 
\begin{equation}
    F[n(\mathbf{r})]
    =
    F[0] 
    +
    \sum_{n=1}^{+\infty} \frac{1}{n!}
    \sum_{\mathbf{G}_1...\mathbf{G}_n} 
    f^{(n)}_{\mathbf{G}_1...\mathbf{G}_n}
    \eta_{\mathbf{G}_1} ... \eta_{\mathbf{G}_n}.
\end{equation}
The crucial point is that the Ginzburg-Landau functional, hence
$f^{(n)}_{\mathbf{r}_1...\mathbf{r}_n}$, need to be invariant under every
symmetry operation of the microscopic lattice.
In particular, translation symmetry by a lattice vector $\mathbf{R}$,
$f^{(n)}_{\mathbf{r}_1 + \mathbf{R}...\mathbf{r}_n + \mathbf{R}}
= f^{(n)}_{\mathbf{r}_1...\mathbf{r}_n}$,
entails 
that one needs
$\mathbf{G}_1+...+\mathbf{G}_n 
\cong 0$ in order to have a nonzero
$f^{(n)}_{\mathbf{G}_1...\mathbf{G}_n}$. 
Here, the symbol $\cong$ denotes 
equality modulo a reciprocal vector with respect to the microscopic lattice;
in this example, this means
$\exp\left\{
i(\mathbf{G}_1+...+\mathbf{G}_n) \mathbf{r}
\right\}=1 \ \forall \mathbf{r}$.

As a corollary, since we are conserving the total particle number, we have that the first order term is zero.
Moreover, assuming a harmonic approximation for which only the lowest reciprocal momenta enter the density expansion, translational invariance entails that for a square crystal the third order term is forbidden, while it is allowed for a triangular crystal.
More precisely, the density in the harmonic approximation reads $n(\mathbf{r})
    = n_0 + \sum_j 
    e^{i\mathbf{G}_j\mathbf{r}}
    \eta_{\mathbf{G}_j}$,
    where for a triangular superlattice 
    $\mathbf{G}_j = \frac{4\pi}{{3} }e^{i\frac{\pi}{3}j}$ with $j=0,1,...,5$ (notice that we used the complex number notation for the 2D vectors), and it is easily noticed that  $\mathbf{G}_0+\mathbf{G}_2+\mathbf{G}_4 \cong 0$
    as well as $\mathbf{G}_1+\mathbf{G}_3+\mathbf{G}_5 \cong 0$. 
    Furthermore, these two allowed terms must enter the free energy expansion with the same weight, because of the (discrete) rotational symmetry 
$f^{(n)}_{e^{i\frac{\pi}{3}}\mathbf{r}_1 ...e^{i\frac{\pi}{3}}\mathbf{r}_n}
= f^{(n)}_{\mathbf{r}_1...\mathbf{r}_n}$.
    With the requirement that both the Ginzburg-Landau functional and the density are real, we arrive at 
\begin{multline}
    F[\eta] = F_0 + 
f_2 \sum\limits_{\mathbf{G}_i}\eta_{\mathbf{G}_i}\eta_{\mathbf{-G}_i} + f_3 \sum\limits_{\mathbf{G}_i + \mathbf{G}_j + \mathbf{G}_k = \mathbf{0}}\eta_{\mathbf{G_i}}\eta_{\mathbf{G}_j}\eta_{\mathbf{G}_k}
+ 
\\ 
+  f_4 \sum\limits_{\mathbf{G}_i + \mathbf{G}_j + \mathbf{G}_k + \mathbf{G}_l = \mathbf{0}}\eta_{\mathbf{G}_i}\eta_{\mathbf{G}_j}\eta_{\mathbf{G}_k}\eta_{\mathbf{G}_l} + ...
    \label{GL_general}
\end{multline}
Using the ansatz $\eta_{\mathbf{G}_j} = \eta > 0$, which entails maximal six-fold rotational symmetry around the density maxima and the that one such maximum is in $\mathbf{r}=0$, one can simplify to 
\begin{equation}
    F[\eta] = F_0 + 
f_2 \eta^2 + f_3 \eta^3 + f_4 \eta^4 + ... 
\end{equation}
where the $f_n$'s are real numbers. A second order phase transition occurs when $f_3=0$ and $f_2$ changes sign. Instead, when $f_3 \neq 0$, a first order phase transition will occur 
when the two local minima cross each other in free energy~\cite{Toledano1987Aug}.

To summarize, the crucial remark is that for a triangular crystal one has $\mathbf{G}_0+\mathbf{G}_2+\mathbf{G}_4 \cong 0$, which is, due to translational invariance, a necessary condition to have the third order term in the free energy. 
Crystallization in a triangular structure will then be generically associated with a first order phase transition. It may occur that $f_3$ is accidentally small, so to have a weakly first order phase transition.

For a square charge density wave, instead, $f_3$ is zero by  symmetry, and a second-order phase transition may occur, even though in practice strong fluctuations and higher order terms can still lead to a first-order transition~\cite{Brazovskii1975,Brazovskii1987}. 
Another example where the $f_3$ has to be zero
is bipartite lattices (e.g. square, honeycomb etc.), whenever the sublattice symmetry is spontaneously broken.

\section{Experimental implementation}
\label{app:exp}

In this Section we try to estimate the relevant range of parameters in which the transition may be observed in TMD experiments.
In the main text we demonstrated that in the extended Hubbard model the IQH-WC transition occurs for $V/t \simeq 3.7$. To relate a realistic heterostructure to these results, we first need to extract the effective Hubbard couplings from a set of experimental parameters.
This can be done by first obtaining the moir\'e bands in the Bistritzer-McDonald continuum approach~\cite{Bistritzer2011}, and then by constructing the Wannier orbitals  of the lowest energy band~\cite{marzari1997,zhang2019bridging}.
For concreteness, we consider 
two kinds of heterostructures:
a MoSe$_2$/WSe$_2$ heterobilayer~\cite{wu2018hubbard}
and a 
WSe$_2$/t-hBN heterostructure.

For the first scenario, the MoSe$_2$/WSe$_2$ heterobilayer, we consider the same parameters as Ref. \cite{moralesduran2022}, i.e $V_M=-11$meV and $\psi = 94^\circ$ to parametrize the moir\'e potential, and take dielectric constant to be $\epsilon=7$.
The experimental knobs are the twist angle $\theta_{\rm twist}$ and the distance from a screening gate $d_{\rm gate}$. The magnetic field needs then to be set so to ensure the proper flux of $2\pi/3$ per moir\'e unit cell.

In Fig.~\ref{fig:WannierHubbard}a) we report 
the ratio $V/t$ calculated in this way, as a function of $\theta_{\rm twist}$ and $d_{\rm gate}$. The (saturated) colorcode has been chosen so that the critical ratio is given by the white regions.
In Fig.~\ref{fig:WannierHubbard}b), for each twist angle we plot the length $a_M$ of the corresponding moir\'e unit cell (solid green line, left axis) and the required magnetic field $B$ (dashed magenta, right axis).
Keeping into account that in standard setups one typically achieves $B \sim 16 T$, a convenient design would consist in choosing
$\theta_{\rm twist} \sim 2^\circ$ and 
$d_{\rm gate} \sim 2$nm, which is definitely accessible experimentally.
Notice that at these high magnetic fields one expects spin-valley polarization~\cite{smolenski2019interaction-induced}, so that one can indeed consider a system of spinless fermions.

The other configuration we consider here is given by a WSe$_2$/t-hBN heterostructure, consisting of a WSe$_2$ monolayer with a proximal twisted hBN.
The monolayer is subject to a purely electrostatic moir\'e potential generated by ferroelectric domains in twisted hBN interfaces.
In a recent experiment,  MoSe$_2$ was put directly in contact with the t-hBN, and a total electrostatic potential depth of $\Delta_{\rm m} \sim 30-40$meV was demonstrated, for a twist angle of $\sim 1.4^\circ$ corresponding to a unit cell of $a_M \sim 10.2$nm~\cite{Kiper2024}. (Here, we propose to use  WSe$_2$ instead of  MoSe$_2$, since the electron effective mass is lighter in the former case, and the strength of electrostatic potential should not significantly change in the two cases.)
In Fig.~\ref{fig:WannierHubbard}c),
we calculated the ratio $V/t$ for electrons in WSe$_2$ as a function of the gate distance and
of $\Delta_{\rm m}$, exploiting the fact that $\Delta_{\rm m}$ can be easily reduced by inserting a (strongly misaligned) hBN spacer between the
WSe$_2$ monolayer and the t-hBN bilayer.
Also, notice that with this moir\'e unit length, the required magnetic flux is achieved for $B \sim 15.3$T.
Therefore, also such structure seems to be extremely promising in order to observe the Hall-Wigner transition.


\section{Conclusion and outlook}
\label{sec:conclusions}

In summary, we have numerically investigated the Integer Quantum Hall to Wigner crystal transition occurring for increasing 
nearest-neighbour repulsion
at one-third density in a triangular lattice subject to an external magnetic
field.
As demonstrated in Sec.~\ref{app:exp}, the critical parameter regime can be experimentally achieved in  twisted MoSe$_2$/WSe$_2$ bilayers
or for a WSe$_2$/t-hBN device.

The evidence for the transition being first-order is consistent with the Landau argument for the free-energy;
sizable fluctuations are nontheless found close to the transition.

It is important to notice that, in continuous systems with screened Coulomb potential and in the absence of magnetic field,
a first-order transition is ruled out by inspection of the surface tension free-energy, which suggests that the liquid-crystal transition occurs via intermediate continuous transitions between bubble and stripe phases~\cite{spivak2004,falson2022competing}. Since we are not aware of any extension of this argument to our lattice setting,  the sharp transition we observe may be a genuine first-order one, and not due to a finite-size
suppression of the bubble and stripe micro-emulsions.

As a theoretical future direction, it will be interesting to investigate the role of spin-spin correlations close to the transition~\cite{moralesduran2023,azadi2024quantum}, as well as the possibility of chiral quantum spin liquid phases on the crystalline side (in analogy to the Mott scenario considered in \cite{kuhlenkamp2022tunable,divic2024chiralspinliquidquantum}).
Development of infinite projected entangled pair states (iPEPS)~\cite{jordan2008classical,orus2009simulation,phien2015infinite} or other tensor network techniques will likely be crucial to tackle the spinful case. 

Finally, it is worth to spend some words concerning the optical detection of the Hall-Wigner transition.
We point out that recent works have studied the formation of umklapp peaks~\cite{smolenski2021signatures,shimazaki2021optical} and analyzed the signatures of the charge gap~\cite{amelio2024edpolaron,amelio2024insulator} in the polaron spectra within the ordered phase. It would be interesting to extend these results in the proximity of the critical point, where the strong quantum fluctuations require the use of powerful numerical methods such as exact diagonalization~\cite{amelio2024edpolaron} or time-dependent tensor network techniques~\cite{vashisht2024chiral}.


\section{Acknowledgments}

We are grateful to 
Michael Knap, Felix Palm and
Laurens Vanderstraeten
  for useful discussions. G.F. was supported by the grant BOF23/GOA/021 from Ghent University and expresses gratitude to the ETH Zurich Solidarity Fund for Foreign Students. I.A. was financially supported by the ERC grant LATIS, the EOS project CHEQS and the FRS-FNRS (Belgium). The work at ETH Zurich was supported by the Swiss National Science Foundation (SNSF) under Grant Number 200021-204076.




\bibliography{biblio.bib}

\end{document}